\documentclass[preprint,showpacs,aps]{revtex4}

\usepackage{epsfig}

\begin{document}

\title{\large \bf Dynamic effect of overhangs and islands at the depinning transition in two-dimensional magnets}

\author{\bf N. J. Zhou and B. Zheng\footnote{corresponding author; email: zheng@zimp.zju.edu.cn}}

\affiliation{Zhejiang University, Zhejiang Institute of Modern
                  Physics, Hangzhou 310027, P.R. China }

\begin{abstract}
With Monte Carlo methods, we systematically investigate the
short-time dynamics of domain-wall motion in the two-dimensional
random-field Ising model with a driving field (DRFIM). We accurately
determine the depinning transition field and critical exponents.
Through two different definitions of the domain interface, we
examine the dynamics of overhangs and islands. At the depinning
transition, the dynamic effect of overhangs and islands reaches
maximum, and this is an important mechanism leading the DRFIM model
to a different universality class from that of the Edwards-Wilkinson
equation with quenched disorder.
\end{abstract}


\maketitle

\section{Introduction}

In the past years, much effort of physicists has been devoted to
domain-wall dynamics in magnetic devices, nanomaterials, thin films,
semiconductors, contact lines, and fluid invasion in porous media
\cite{he92,lem98,ono99,mou04,yam07,im09}. In particular, quenched
randomness in ferroic materials, e.g. ultrathin ferromagnetic and
ferroelectric films, fundamentally affects the response to an
external field \cite{met07,shi07,dou08,kim09}. From a pragmatical
point of view, understanding the controlled movement of domain walls
plays an important role in developing new classes of potential
non-volatile storage-class memories \cite{shi07,hay08,par08}. From a
purely theoretical point of view, it is also essential for
understanding non-equilibrium dynamics in disordered media
\cite{kol05a,kol06,kol09,zho09}. For a dc (direct current) driving
field, $H$, the domain-wall motion exhibits a depinning transition
at zero temperature. The depinning field, $H_c$, separates the
regimes of static pinning ($H < H_c$) and friction-limited viscous
slide ($H > H_c$) \cite{now98,due05,kol06a,bak08}. At low, but
non-zero, temperatures, the sharp depinning transition is softened
and a thermally activated creep state appears \cite{rot01,kol05}.
For an ac (alternative current) driving field, $H(t)=H_0 exp(i2\pi
ft)$, and at a nonzero temperature, the domain-wall motion exhibits
different states and dynamic phase transitions, which can be
classified in the so-called Cole-Cole diagrams
\cite{bra05,kle07,jez08}.

Current theoretical approaches to domain-wall dynamics in ultrathin
films are typically based on the Edwards-Wilkinson equation with
quenched disorder (QEW) \cite{nat92,due05,kol06,bus08}. This
equation is a phenomenological model, and detailed microscopic
structures and interactions of real materials are not concerned.
Additionally, a self-inconsistence is puzzling at the depinning
transition, especially for the roughness exponent $\zeta$
\cite{ros01,les93, jen95}. To understand the domain-wall motion at a
microscopic level, one needs lattice models based on microscopic
structures and interactions \cite{now98,rot01,col06}. The
random-field Ising model with a driving field (DRFIM) is an
candidate, at least to capture robust features of the domain-wall
motion, although it does not include all interactions in real
materials.

As an interface propagates, overhangs and islands may occur. Such a
phenomenon is observed in many experiments for magnetic materials
\cite{lee09,jos98}. In the QEW equation, an interface is described
by a single-valued elastic string, and overhangs and islands are not
taken into account \cite{kol05,jos96,ros03}. However, the anomalous
roughness exponent ($\zeta \neq 1$) has led ones to conjecture that
an one-dimensional string necessarily develops overhangs and islands
\cite{ros01}. In the DRFIM model, overhangs and islands are
naturally created, thus the domain wall is not single-valued and
one-dimensional \cite{urb95,ama95,rot00}. Numerical simulations of
the DRFIM model show that characteristics of the domain-wall motion
may depend on overhangs and islands \cite{rot99,zho09}. In the
literatures, however, the dynamics of overhangs and islands is
rarely referred, and its dynamic effect on the depinning transition
is not identified.

In the past years much progress has been achieved in critical
dynamics far from equilibrium \cite{jan89,hus89,zhe99,zhe98}.
Although the spatial correlation length is still short in the
beginning of the time evolution, the short-time dynamic scaling form
is induced by the divergent correlating time around a continuous
phase transition. Based on the {\it short-time} dynamic scaling
form, new methods for the determination of both dynamic and static
critical exponents have been developed \cite{zhe98,luo98}. Since the
measurements are carried out in the short-time regime, one does not
suffer from critical slowing down. Recent activities include various
applications and developments such as theoretical and numerical
studies of the Josephson junction arrays and aging phenomena
\cite{gra05,cal05,lei07,lin08,lee05}. A kind of domain-wall
roughening process at order-disorder phase transitions has also been
revealed \cite{zho07,zho08,he09}. Very recently, the short-time
behavior of the domain-wall relaxation around the depinning
transition is noted in simulations and experiments
\cite{rod07,kol06,kol09}.

In a recent article \cite{zho09}, based on the short-time dynamic
approach, the domain-wall dynamics of a disordered magnetic system
driven by a constant field $H$ is investigated with the DRFIM model,
in comparison with the QEW equation and experiments. The depinning
transition at zero temperature is of second order, and its ordered
parameter is the interface velocity. The transition field, static
and dynamic exponents, and local and global roughness exponents are
accurately determined for the DRFIM model. The results indicate that
the DRFIM model does not belong to the universality class of the QEW
equation \cite{zho09}, in contrast to the usual assumption
\cite{ros03,ama95}.

However, it is unknown what mechanism leads the DRFIM model to a
different universality class from the QEW equation. The fluctuation
and correlation of the interface velocity, and especially, the
dynamics of overhangs and islands are not touched at all. Due to the
existence of overhangs and islands, the definition of the domain
interface is theoretically not unique, and comparison with
experiments remains ambiguous. The purpose of this paper is to
provide a comprehensive understanding of the depinning transition in
the DRFIM model. We emphasize that the dynamics of overhangs and
islands plays a key role. In Sec. II, the model and scaling analysis
are described. In Sec. III, numerical results are presented. Sec. IV
includes the conclusions.

\section{Model and scaling analysis}

\subsection{Model}

 The random-field Ising model is defined by the Hamiltonian
\begin{equation}
\mathcal{H} = - J \sum_{<ij>}S_iS_j - H \sum_i S_i - \sum_i h_i S_i,
\label{equ10}
\end{equation}
where $S_i = \pm 1$ is the Ising spin of the two-dimensional square
lattice. The random field $h_i$ is uniformly distributed within an
interval $[-\Delta, \Delta]$, and $H$ is a homogeneous driving
field. Following Refs.~\cite{now98,zho09}, we fix $\Delta = 1.5 J$
and set $J=1$. For comparison, $\Delta = 0 $ is also simulated. A
Gaussian distribution of the random field $h_i$ leads to similar
results, but it is technically more complicated. Therefore, the
results of the uniform distribution of $h_i$ are presented in this
paper. Our simulations are performed at zero temperature with
lattice sizes $L=128, 256, 512$, and $1024$ up to $t_{max}=2000$,
with total samples $10,000, 40,000, 50,000$, and $30,000$,
respectively. Simulations of different $L$ confirm that our results
do not suffer from finite-size effects. Errors are estimated by
dividing the samples into three or four subgroups. If the
fluctuation of the curve in the time direction is comparable with or
larger than the statistical error, it will be taken into account.

The initial state is such a state, that spins are positive in the
sublattice on the left side and negative on the right side. We set
the $x$ axis in the direction perpendicular to the perfect domain
wall. Antiperiodic and periodic boundary conditions are used in $x$
and $y$ directions, respectively. To eliminate the pinning effect
irrelevant for disorder, we rotate the square lattice such that the
initial domain wall orients in the $(11)$ direction of the square
lattice, as shown in Refs.~\cite{now98,rot99,rot01,zho09}. After
preparing the initial state, we {\it randomly} select a spin, and
flip it if the total energy decreases after flipping. A Monte Carlo
time step is defined by $L^2$ single-spin selects. As time evolves,
the domain wall moves and roughens while the bulk remains unchanged.

We should emphasize that overhangs and islands are naturally created
in the DRFIM model during the time evolution, but only {\it by the
domain wall}, since the temperature is set to zero. Therefore, it
makes sense to define the domain wall as an interface, the so-called
domain interface. Due to the existence of overhangs and islands,
however, there may be different ways to define the domain interface.
In this paper, two typical definitions of the domain interface are
concerned, to investigate the dynamics of overhangs and islands, and
to compare with the QEW equation and experiments. We first study the
definition commonly used in the literatures, i.e., the one define by
the magnetization \cite{now98,rot01,zho09}. Denoting a spin at site
$(x, y)$ by $S_{xy}(t)$, we introduce a {\it line} magnetization
\begin{equation}
m(y,t) = \frac{1}{L} \left[ \sum_{x=1}^L S_{xy}(t) \right].
 \label{equ20}
\end{equation}
The height function of the domain interface is then defined as
\begin{equation}
h_m(y,t) = \frac{L}{2}[ m(y,t) + 1] . \label{equ30}
\end{equation}
The subscript $m$ indicates that it is defined by the line
magnetization. Thus the roughness function is introduced to depict
the roughening of the domain interface,
\begin{equation}
\omega_m^{(2)}(t) = \left \langle h_m(y,t)^2 \right \rangle -
\langle h_m(y,t) \rangle^2, \label{equ40}
\end{equation}
where $<\cdots>$ includes the statistical average and average over
$y$. A more informative quantity is the height correlation function
\cite{jos96},
\begin{equation}
C_m(r, t) = \left\langle[h_m(y + r, t) - h_m(y, t)]^2 \right
\rangle. \label{equ50}
\end{equation}
It describes both the spatial correlation of the height function in
the $y$ direction and the growth of the domain interface in the $x$
direction. To {\it independently} estimate the dynamic exponent $z$,
we introduce an observable
\begin{equation}
F_m(t) = [M^{(2)}(t) - M(t)^2]/\omega_m^2(t). \label{equ55}
\end{equation}
Here $M(t)$ is the {\it global} magnetization and $M^{(2)}(t)$ is
its second moment. In fact, $F_m(t)$ is nothing but the ratio of the
planar susceptibility and line susceptibility.

On the other hand, the local interface velocity is defined as the
time derivative of the height function \cite{now98}
\begin{equation}
v_m(y,t) = \frac{dh_m(y,t)}{dt}. \label{equ60}
\end{equation}
The average velocity of the domain interface is then obtained,
\begin{equation}
v_{_M}(t) = v_{m}(t)=\langle v_m(y,t) \rangle. \label{equ70}
\end{equation}
Here we use the subscript $M$ to emphasize that $v_{_M}(t)$ is the
global average velocity. In fact, $v_{_M}(t)$ is the order parameter
of the depinning phase transition. The local and global fluctuations
of the local interface velocity, $v_m^{(2)}(t)$ and
$v_{_M}^{(2)}(t)$, are interesting observables
\begin{equation}
v_m^{(2)}(t) =  \langle  v_m(y,t)^2 \rangle - v_{_M}(t)^2,
\label{equ80}
\end{equation}
\begin{equation}
v_{_M}^{(2)}(t) = \left \langle \left[ \frac{1}{L} \sum_{y=1}^L
v_m(y,t)\right]^2 \right \rangle - v_{_M}(t)^2. \label{equ90}
\end{equation}
In Eq.~(\ref{equ90}), $<\cdots>$ only includes the statistical
average. We note that $v_m^{(2)}(t)$ describes the fluctuation of
the interface velocity in the $x$ direction, while $v_{_M}^{(2)}(t)$
also includes the correlation of the interface velocity in the $y$
direction.

\begin{figure}[ht]
\epsfysize=7.0cm \epsfclipoff \fboxsep=0pt
\setlength{\unitlength}{1.cm}
\begin{picture}(10,6)(0,0)
\put(1.0,0.3){{\epsffile{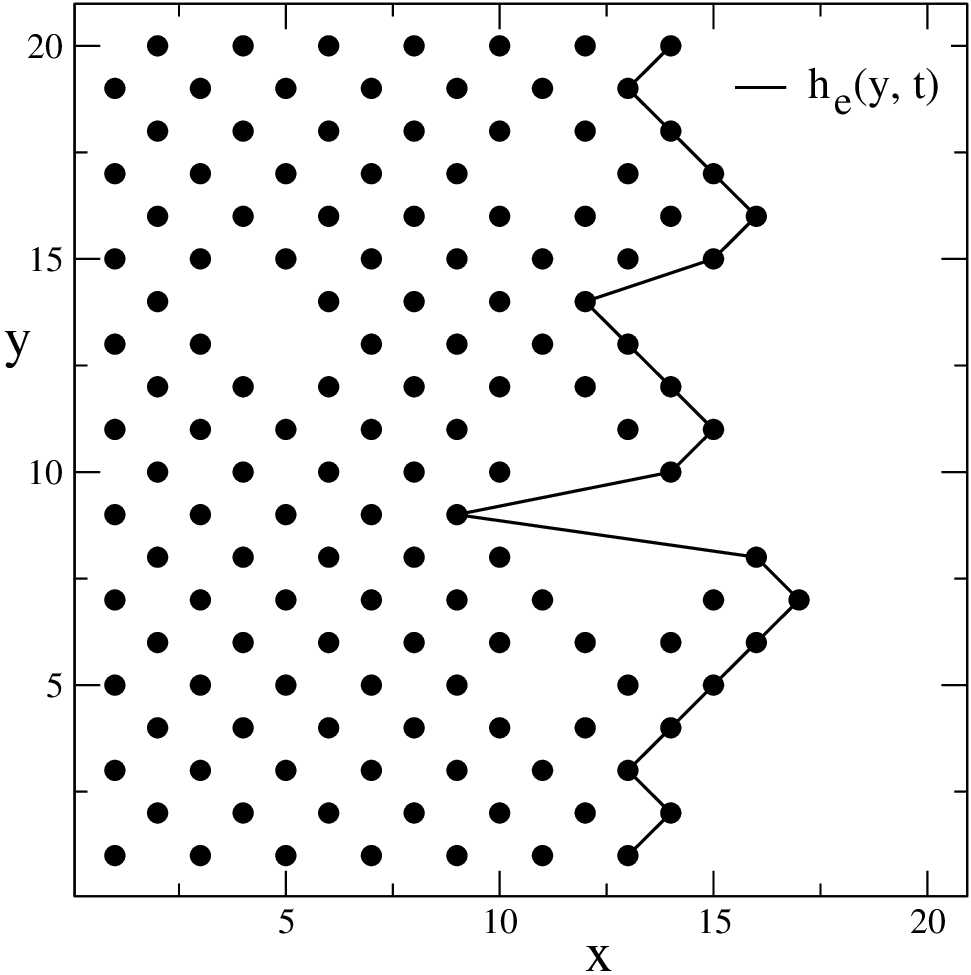}}}
\end{picture}

\caption{The height function $h_e(y,t)$ is defined as the envelop of
the positive spins.}\label{f0}
\end{figure}

In the definition of the height function in Eq.~(\ref{equ30}),
overhangs and islands looks formally suppressed. However, they {\it
do affect} the dynamic evolution of the spin configuration, and the
dynamic effect of overhangs and islands is partially included in the
interface propagation and growth \cite{zho09}. Due to the existence
of overhangs and islands, however, the definition of the domain
interface is not unique. Whether different definitions lead to the
same results, and how to compare with the QEW equation and
experiments, remain ambiguous. For example, another definition of
the height function can be introduced by the {\it envelop} of the
positive spins, as shown in Fig.~\ref{f0}. In other words, we define
the height function $h_e(y,t)$ as the largest $x$ coordinate of the
positive spins with the fixed $y$ coordinate and at the time $t$.
This definition is closer to the experiments with imaging technique
\cite{lem98,lee09,jos98}.

With $h_e(y,t)$, one can derive the local interface velocity
$v_e(y,t)$. Similar to Eqs.~(\ref{equ40})-(\ref{equ90}), the average
velocity $v_{_E}(t)$, the roughness function $\omega_{e}^2(t)$, the
function $F_e(t)$, the height correlation function $C_{e}(r,t)$, and
the local and global fluctuations of the velocity, $v_e^{(2)}(t)$
and $v_{_E}^{(2)}(t)$, can be calculated. In the definition of the
height function with the envelop, the dynamic effect of overhangs
and islands is {\it maximally} taken into account. In contrast to
it, the height function defined with the magnetization includes only
a minimal contribution of overhangs and islands.

To reveal the dynamic characteristics of overhangs and island,
therefore, we may introduce the overhang height function $\delta
h(y,t)$ and local overhang velocity $\delta v(y,t)$
\begin{equation}
\delta h(y,t) = h_{e}(y,t) - h_{m}(y,t), \label{equ95}
\end{equation}
\begin{equation}
\delta v(y,t) = v_{e}(y,t) - v_{m}(y,t). \label{equ97}
\end{equation}
Then the average overhang velocity is
\begin{equation}
\delta v(t) = v_{_E}(t) - v_{_M}(t), \label{equ100}
\end{equation}
and it is related to the average overhang size by
\begin{equation}
\delta h(t) = \int_0^t \delta v(t') dt'. \label{equ110}
\end{equation}
Similar to Eqs.~(\ref{equ40}), (\ref{equ80}) and (\ref{equ90}), the
roughness function of overhangs and islands, $\delta h^{(2)}(t)$,
the local and global fluctuations of the overhang velocity, $\delta
v_l^{(2)}(t)$ and $\delta v_G^{(2)}(t)$, can be defined,
\begin{equation}
\delta h^{(2)}(t) = \langle [\delta h(y,t)]^2 \rangle - [\delta
h(t)]^2, \label{equ120}
\end{equation}
\begin{equation}
\delta v_l^{(2)}(t) = \langle  [\delta v(y,t))]^2 \rangle - [\delta
v(t)]^2, \label{equ130}
\end{equation}
\begin{equation}
\delta v_G^{(2)}(t) = \left \langle  \left[\frac{1}{L} \sum_{y=1}^L
\delta v(y,t)\right]^2 \right \rangle - [\delta v(t)]^2.
\label{equ140}
\end{equation}
To measure the dynamic exponent independently, we can construct a
function $F_\delta (t)$ in a similar form of Eq.~(\ref{equ55}).

\subsection{Scaling analysis}
Since the depinning transition is a second-order phase transition,
the dynamic evolution of the order parameter $v_{_M}(t)$ should obey
the dynamic scaling theory supported by the renormalization-group
calculations \cite{jan89,zhe98,luo98}. For a finite lattice size
$L$, and assuming a nonequilibrium correlation length $\xi \sim
t^{1/z}$, scaling arguments lead to a dynamic scaling form for the
order parameter \cite{jan89,zhe98,luo98},
\begin{equation}
v_{_M}(t, \tau, L) = b^{-\beta/\nu} G(b^{-z}t, b^{1/\nu}\tau,
b^{-1}L). \label{equ150}
\end{equation}
Here $b$ is an arbitrary rescaling factor, $\beta$ and $\nu$ are the
static exponents, $z$ is the dynamic exponent, and
$\tau=(H-H_c)/H_c$. Setting $b \sim \xi(t) \sim t^{1/z}$, the
dynamic scaling form is rewritten as
\begin{equation}
v_{_M}(t, \tau, L) = t^{-\beta/\nu z} G(1, t^{1/\nu z}\tau,
t^{-1/z}L). \label{equ160}
\end{equation}
In the short-time regime, i.e., the regime with  $\xi(t)\sim t^{1/z}
\ll L$, the finite-size effect is negligibly small,
\begin{equation}
v_{_M}(t, \tau) = t^{-\beta / \nu z}G(t^{1/\nu z}\tau).
\label{equ170}
\end{equation}
Therefore, at the transition point $\tau = 0$, a power law behavior
is obtained,
\begin{equation}
v_{_M}(t) = t^{-\beta / \nu z}. \label{equ180}
\end{equation}
With Eq.~(\ref{equ170}), the critical field $H_c$ may be located by
searching for the best power-law behavior of $v_{_M}(t, \tau)$
\cite{zhe98,luo98}. The critical exponent $\beta /\nu z$ is then
estimated from Eq.~(\ref{equ180}). The critical exponent $1/\nu z$
can be obtained from the time derivative of $v_{_M}(t, \tau)$,
calculated according to Eq.~(\ref{equ170}) \cite{zho09}.

In general, even at the transition point, the roughness function
$\omega_m^2(t)$, the height correlation function $C_m(r,t)$ and the
global fluctuation $v_{_M}^{(2)}(t)$ of the interface velocity in
Eqs. (\ref{equ40}), (\ref{equ50}) and (\ref{equ90}), may not obey a
perfect power-law behavior in early times. In fact, the domain
interface and its velocity also roughen {\it even without disorder}
($\Delta=0$), due to the random updating scheme in numerical
simulations. This may induce corrections to scaling. To capture the
dynamic effect of disorder, we introduce the pure roughness function
$D\omega_m^2(t)$, the pure height correlation function $DC_m(r, t)$,
and the pure global fluctuation $Dv_{_M}^{(2)}(t)$ by subtracting
the contributions of $\Delta=0$, $\omega_{m,b}^2(t)$,
$C_{m,b}(r,t)$, and $v_{_M, b}^{(2)}(t)$, respectively. For a
sufficiently large lattice and at the transition point, we should
observe standard power-law scaling behaviors
\cite{jos96,zho08,bak08,zho09},
\begin{equation}
D\omega_m^2(t) \sim t^{ 2 \zeta / z}, \label{equ190}
\end{equation}
and
\begin{equation}
   DC_m(r,t) \sim  \left\{
   \begin{array}{lll}
     t^{2(\zeta-\zeta_{loc})/z}\ r^{2\zeta_{loc}}   & \quad &  \mbox{if  $r \ll \xi(t) \ll L$} \\
     t^{2\zeta /z}  & \quad &  \mbox{if $0 \ll \xi(t) \ll r$}
   \end{array}\right. .
   \label{equ200}
\end{equation}
Here $\xi(t) \sim t^{1/z}$, $\zeta$ is the global roughness
exponent, and $\zeta_{loc}$ is the local one. Finally, the dynamic
exponent $z$ is independently determined by
\begin{equation}
F_m(t) \sim t^{1/z}/L. \label{equ205}
\end{equation}

Compared with the velocity itself, the global and local fluctuations
of the velocity exhibit more complicated dynamic behaviors. In fact,
there are {\it two intrinsic length scales} at the depinning
transition, the correlation length $\xi(t) \sim t^{1/z}$ in the $y$
direction, and the characteristic length of roughening in the $x$
direction, $l(t)\sim t^{\zeta / z}$, defined in Eq.~(\ref{equ190}).
It only happens that $l(t)$ is irrelevant for the average velocity
$v_{_M}(t)$. In general, therefore, the pure global fluctuation of
the velocity should obey the scaling form \cite{zhe98,luo98}
\begin{equation}
Dv_{_M}^{(2)}(t) = t^{-2\beta / \nu z}F(\xi(t)/L, l(t)/\xi(t)).
\label{equ210}
\end{equation}
For a sufficiently large lattice, finite-size scaling analysis leads
to \cite{zhe98,luo98}
\begin{equation}
Dv_{_M}^{(2)}(t) \sim (\xi(t)/L)^d , \label{equ220}
\end{equation}
where $d = 1$ is the spatial dimension of the domain interface. Our
numerical simulations show that the dependence of $Dv_{_M}^{(2)}(t)$
on $l(t)/\xi(t)$ takes also a power law. Therefore,
\begin{equation}
Dv_{_M}^{(2)}(t) \sim t^{ [1 - 2 \beta / \nu  + \lambda_{_M}(\zeta -
1)]/z }/L, \label{equ230}
\end{equation}
where $\lambda_{_M}$ is an exponent, reflecting the dynamic effect
of the roughness of the domain interface. Similarly, the local
fluctuation of the velocity corresponds to $d = 0$,
\begin{equation}
v_m^{(2)}(t) \sim t^{[- 2 \beta / \nu + \lambda_m(\zeta - 1)]/z}.
\label{equ240}
\end{equation}
In fact, our numerical simulations yield $\lambda_{_M} \approx 2$
and $\lambda_{m} \approx 3$. For the standard order-disorder phase
transition, such terms described by $\lambda_{_M}$ and $\lambda_{m}$
do not exist, since the roughness exponent $\zeta=1$
\cite{zho08,he09}. This is a difference between the depinning
transition and the standard order-disorder phase transition.

Due to the dynamic effect of overhangs and islands, it is unclear
whether the domain interface defined by the envelop of the positive
spins also obeys the dynamic scaling forms described in Eqs.
(\ref{equ150})-(\ref{equ240}). In fact, we may also examine the
dynamics of overhangs and islands independently. Similar to the
determination of the critical field $H_c$ from Eq.~(\ref{equ170}),
one can locate the critical field $H_c$ by searching for the best
power-law behavior of the overhangs velocity $\delta v(t, H)$. At
the critical point $H_c$, however, the overhang velocity $\delta
v(t)$ {\it increases} with time, different from the interface
velocity $v(t)$. For example, we may write
\begin{equation}
\delta v / v \sim t^\theta. \label{equ250}
\end{equation}
Here $\theta$ is estimated to be about $0.5$ in our numerical
simulations. In general, $\delta h(y,t)$ and $\delta v(y,t)$ are
dynamic variables independent of $h_m(y,t)$ and $v_m(y,t)$, and one
needs another set of critical exponents to describe their dynamic
behaviors. Similar to $D\omega^{(2)}(t)$ in Eq.~(\ref{equ190}), the
pure roughness function of overhangs and islands obeys
\cite{zho09,zho08,he09}
\begin{equation}
D\delta h^{(2)}(t) \sim t^{2\zeta_{\delta} / z_{\delta}},
\label{equ260}
\end{equation}
where $\zeta_{\delta}$ and $z_{\delta}$ are the roughness exponent
and dynamic exponent of overhangs and islands, respectively. The
dynamic exponent $z_{\delta}$ can be independently determined from
\begin{equation}
F_\delta(t) \sim t^{1/z_{\delta}}/L. \label{equ265}
\end{equation}
Since the overhang velocity $\delta v(y, t)$ does not show
correlation in the $y$ direction, we assume that both the local and
global fluctuations of the overhang velocity, $\delta v_l^{(2)}(t)$
and $\delta v_G^{(2)}(t)$ defined in Eqs.~(\ref{equ130}) and
(\ref{equ140}) obey
\begin{equation}
\delta v^{(2)}(t) \sim t^{2\alpha}. \label{equ270}
\end{equation}
On the other hand, the global fluctuation $\delta v_G^{(2)} \sim
L^{-1}$, while the local one $\delta v_l^{(2)}$ and the roughness
function $\delta h^{(2)}(t)$ are $L$-independent.

\section{Numerical simulations}

In Ref. \cite{zho09}, the dynamic scaling behaviors from
Eq.~(\ref{equ170}) up to Eq.~(\ref{equ205}) have been carefully
examined for the domain interface defined with the magnetization in
the DRFIM model. The relevant critical exponents are accurately
determined, and are summarized in Table~\ref{t1}, in comparison with
those of the QEW equation. Our first task in this paper is to
investigate whether these dynamic scaling forms hold also for the
domain interface defined with the envelop of the positive spins.

\begin{figure}[ht]
\epsfysize=7.0cm \epsfclipoff \fboxsep=0pt
\setlength{\unitlength}{1.cm}
\begin{picture}(10,6)(0,0)
\put(-3.0,-0.3){{\epsffile{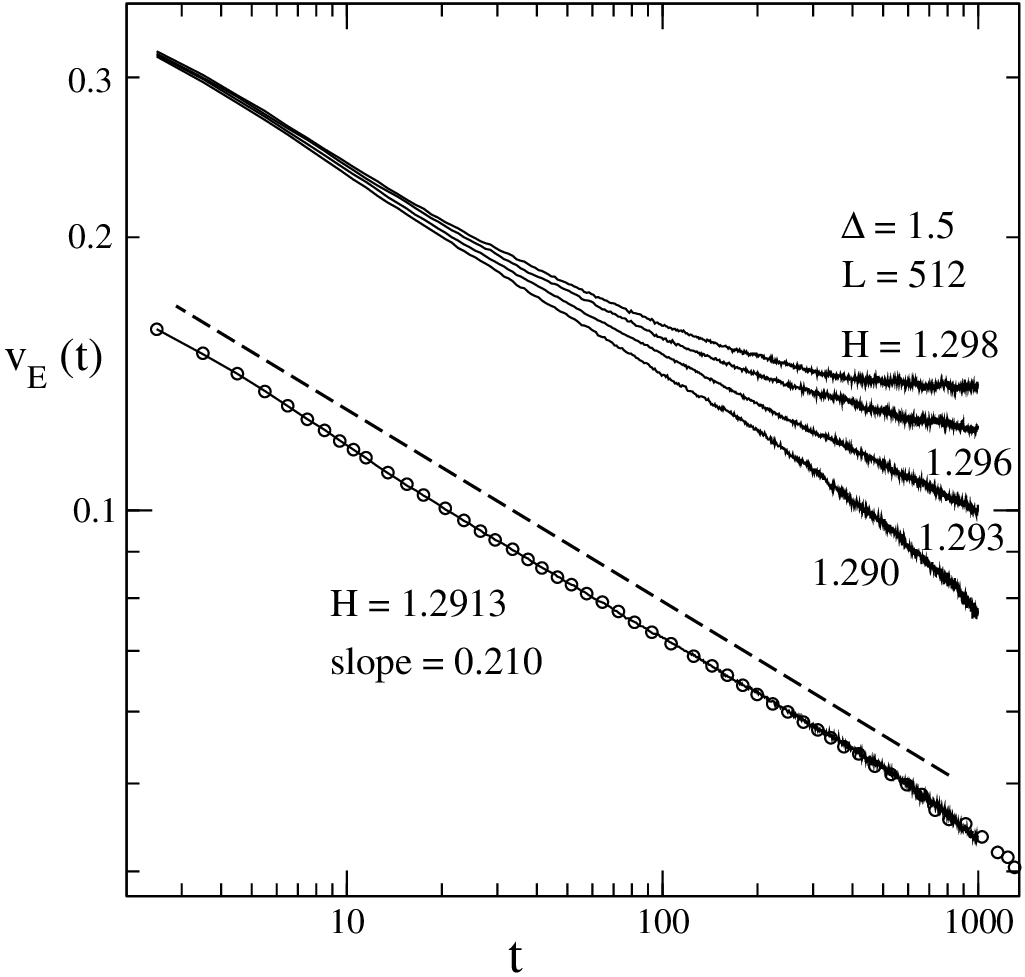}}}\epsfysize=7.0cm
\put(5.2,-0.3){{\epsffile{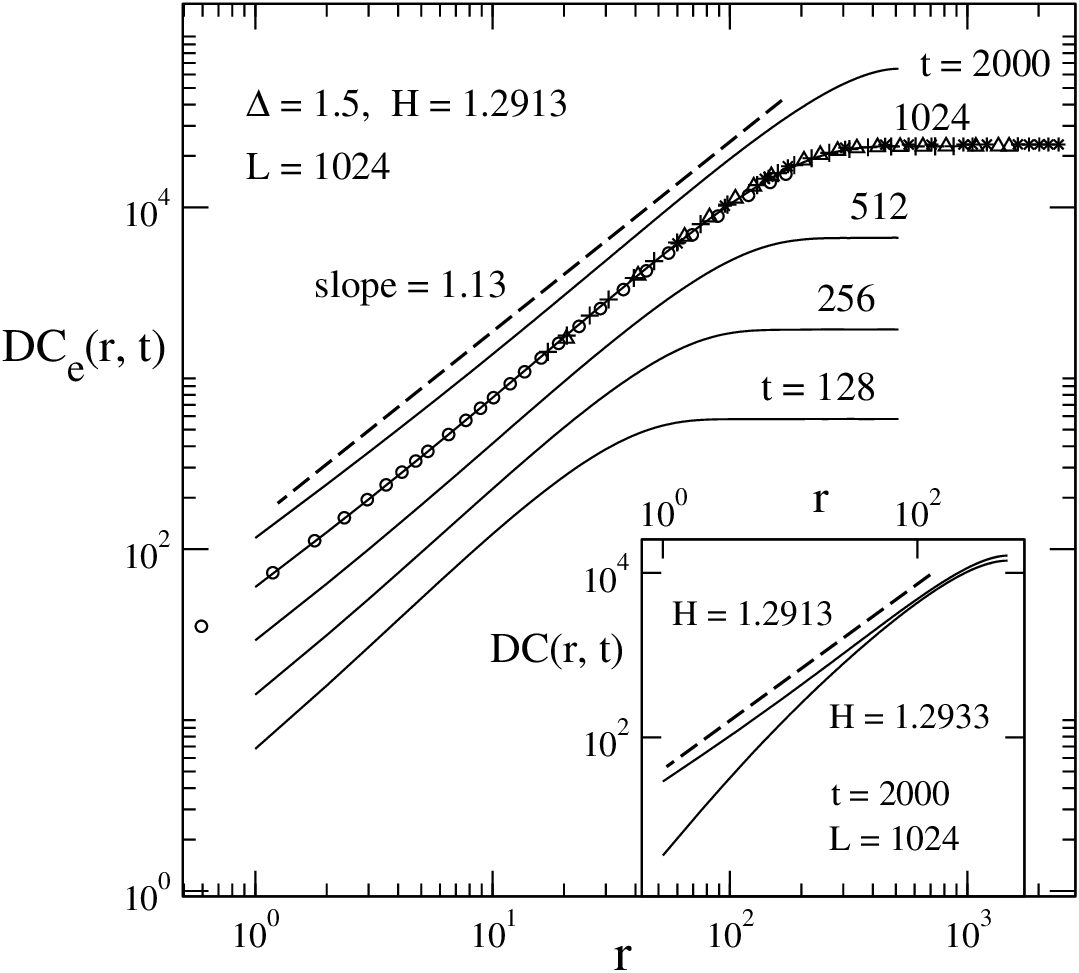}}}
\end{picture}

\hspace{-5.0cm}\footnotesize{(a)}\hspace{8.0cm}\footnotesize{(b)}
\caption{(a) Interface velocity $v_{_E}(t, \tau)$ is plotted with
solid lines for different driving fields $H$ with $L=512$ on a
log-log scale. For clarity, the curve of $H = 1.2913$ is shifted
down. For comparison, the curve with $L = 1024$ is shown with open
circles. (b) The pure height correlation function $DC_{e}(r, t)$ is
displayed at $H_c = 1.2913$. According to Eq.~(\ref{equ220}), data
collapse is demonstrated. Stars, triangles, pluses, and circles
correspond to $t = 128, 256, 512$, and $2000$, respectively. In the
inset, $DC(r,t)$ defined with the envelop (upper) and with the
magnetization (lower) are shown at $t=2000$. In both (a) and (b),
dashed lines show power-law fits. }\label{f1}
\end{figure}

In Fig.~\ref{f1}(a), the average interface velocity $v_{_E}(t,
\tau)$ is displayed for different driving field $H$. It drops
rapidly down for smaller $H$, while approaches a constant for larger
$H$. Due to the dynamic effect of overhangs and islands, $v_{_E}(t,
\tau)$ does {\it not} exhibit a power-law scaling behavior at the
transition field $H_c=1.2933$ located from $v_{_M}(t, \tau)$. By
searching for the best power-law behavior of $v_{_E}(t,\tau)$,
however, one could detect an alternative transition field $H_c =
1.2913(4)$, which is slightly smaller than $H_c = 1.2933(2)$
obtained from $v_{_M}(t, \tau)$. Although the difference looks
small, it is not a statistical error. We believe that this
difference is due to the dynamic effect of overhangs and islands at
the non-stationary stage of the dynamic evolution. Anyway, the
critical exponent $\beta / \nu z = 0.210(2)$ is obtained from the
slope of the curve at $H_c$, according to Eq.~(\ref{equ180}). In a
similar way, we measure $1/\nu z = 0.76(3)$, $2\zeta / z = 1.78(1)$
and $1/ z = 0.778(7)$, based on Eqs.~(\ref{equ170}), (\ref{equ190})
and (\ref{equ205}) respectively \cite{zho09}. We then calculate the
critical exponents $\beta =0.278(4)$, $\nu =1.02(4)$, $z=1.28(1)$
and $\zeta =1.14(1)$.

In Fig.~\ref{f1}(b), the pure height correlation function
$DC_{e}(r,t)$ is displayed as a function of $r$ at $H_c$ for
different time $t$ . According to Eq.~(\ref{equ200}), the critical
exponent $2 \zeta_{loc} = 1.13(2)$ is derived from the slope of the
curve at a large time $t = 2000$. To fully confirm the scaling form
of $DC_{e}(r,t)$ in Eq.~(\ref{equ200}), for example, we fix $t' =
1024$, and rescale $r$ of another $t$ to $(t'/t)^{1/z}r$, and
$DC_{e}(r,t)$ to $(t'/t)^{2\zeta/z}DC_{e}(r,t)$. As shown in
Fig.~\ref{f1}(b), data of different $t$ nicely collapse to the curve
of $t' = 1024$ with $\zeta = 1.14$ and $z = 1.28$ as input. Hence
the scaling form is validated. In experiments, the critical exponent
$\zeta_{loc}$ is usually measured from the power-law behavior
$DC_{e}(r,t) \sim r^{2 \zeta_{loc}}$ at a large $t$. Indeed, this
power-law behavior of $DC_{e}(r,t)$ is much cleaner than that of
$DC_{m}(r,t)$, as shown in the inset. Plotting $DC_{e}(r,t)$ as a
function of $t$ for different $r$, we measure $2\zeta / z = 1.78(1)$
and $2(\zeta - \zeta_{loc}) / z = 0.890(5)$, based on
Eq.~(\ref{equ200}). With the dynamic exponent $z=1.28(1)$ as input,
we may calculate $\zeta$ and $\zeta_{loc}$. The global roughness
exponent $\zeta =1.14(1)$ is the same as that obtained from the
roughness function, and the local roughness exponent $\zeta_{loc}
=0.569(6)$ is consistent with $\zeta_{loc} = 0.565(10)$ measured
directly in Fig.~\ref{f1}(b).

All the measurements of the critical exponents and transition field
are summarized in Table~\ref{t1}, in comparison with those for the
domain interface defined with the magnetization and for the QEW
equation. These results further confirm that the DRFIM model and QEW
equation are not in a same universality class \cite{zho09,urb95}.
For the domain interface defined with the magnetization, the
exponents $\beta, z$ and $\zeta$ of the DRFIM model differ from
those of the QEW equation by about $10$ percent, and especially, the
difference of $\nu$ and $\zeta_{loc}$ between two models reaches
nearly $30$ percent \cite{zho09}. For the domain interface defined
with the envelop, the difference is even larger. For the exponent
$\zeta_{loc}$, for example, the difference between two models is
about $45$ percent. These results suggest that it is mainly the
overhangs and islands that induces the difference between the DRFIM
model and QEW equation. Although real materials may include more
complicated interactions than the DRFIM model, experimental
measurements of the local roughness exponent of the domain interface
support $\zeta_{loc}<1$. For $T>0$ and $0<H<H_c$, for example, it is
reported that $\zeta_{loc}=0.7(1)$ and $0.69(7)$ in the experiments
with ultrathin Pt/Co/Pt films \cite{met07,lem98}, and
$\zeta_{loc}=0.78(1)$ with Co$_{28}$Pt$_{72}$ alloy films
\cite{jos98}.

\begin{figure}[ht]
\epsfysize=7.0cm \epsfclipoff \fboxsep=0pt
\setlength{\unitlength}{1.cm}
\begin{picture}(10,6)(0,0)
\put(-3.0,-0.3){{\epsffile{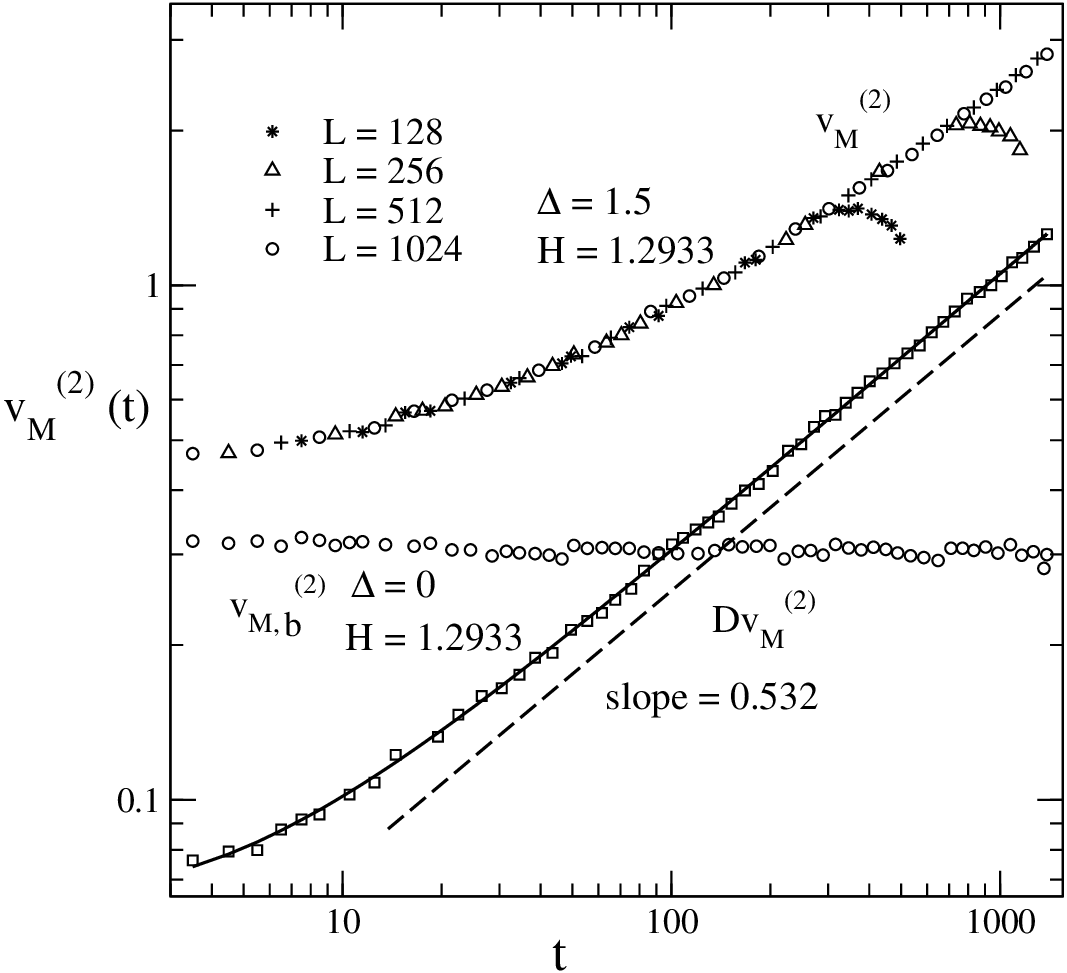}}}\epsfysize=7.0cm
\put(5.2,-0.3){{\epsffile{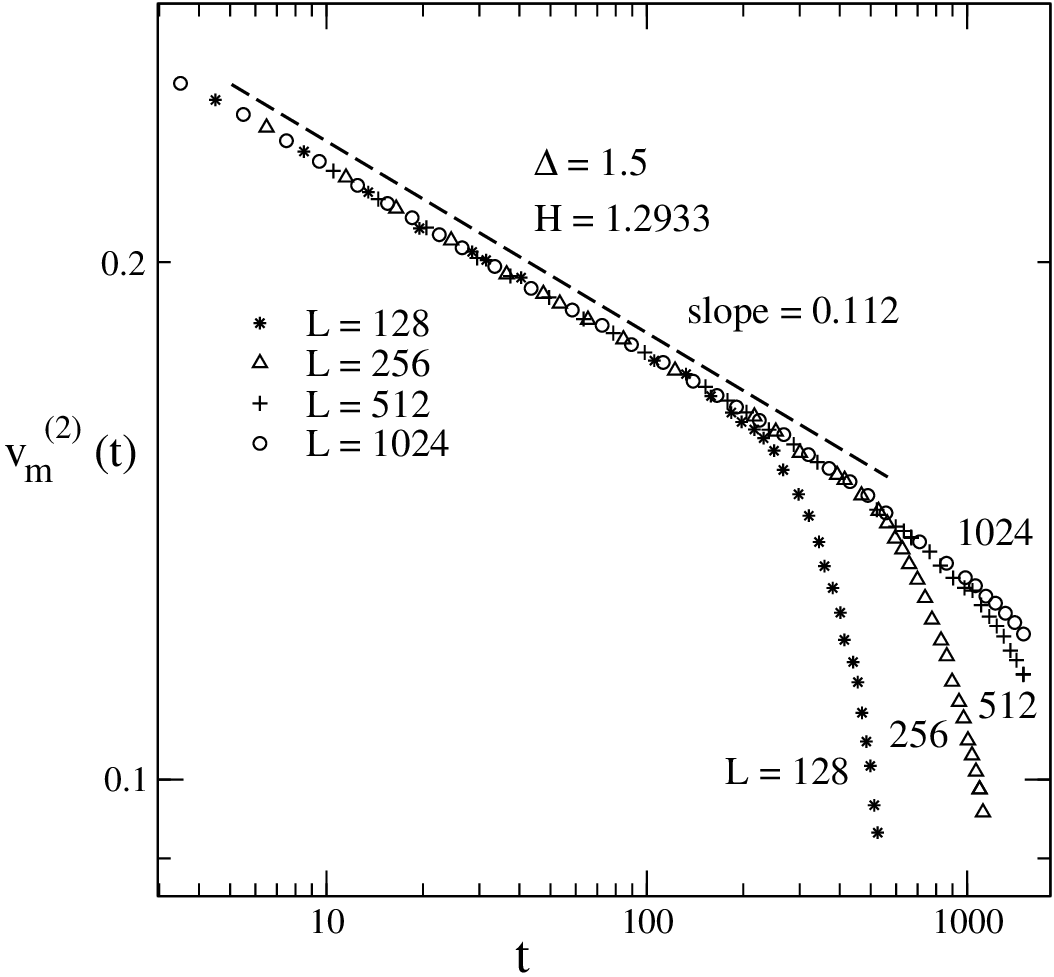}}}
\end{picture}

\hspace{-5.0cm}\footnotesize{(a)}\hspace{8.0cm}\footnotesize{(b)}
\caption{In (a) and (b), the global and local fluctuations of the
interface velocity are displayed at $H_c = 1.2933$ for different $L$
on a log-log scale. Dashed lines show power-law fits, and the solid
line represents a power-law fit with correction. In (a), the pure
global fluctuation function $Dv_{_M}^{(2)}(t)$ is shown for $L =
1024$ with squares. All the curves have been rescaled by a factor
$L$. }\label{f2}
\end{figure}

In the stationary state, the depinning transition should be uniquely
defined. Therefore, the transition field located from the short-time
dynamic behavior of $v_{_E}(t, \tau)$ is an effective one. In other
words, $v_{_E}(t, \tau)$ is not a "good" order parameter in rigorous
sense, although it is closer to the experiments with imaging
technique \cite{lem98,lee09,jos98}. To demonstrate the difference
between the domain interfaces defined with the magnetization and
envelop, we may consider the fluctuation of the interface velocity.
In Fig.~\ref{f2}(a), the global fluctuations of the interface
velocity, $v_{_M}^{(2)}(t)$ and $v_{_M,b}^{(2)}(t)$ for $\Delta =
1.5$ and $0$, and the pure global fluctuation $Dv_{_M}^{(2)}(t) =
v_{_M}^{(2)}(t) - v_{_M,b}^{(2)}(t)$ are displayed. Obviously
$Dv_{_M}^{(2)}(t)$ shows a cleaner power-law behavior than
$v_{_M}^{(2)}(t)$ does, due to the subtraction of
$v_{_M,b}^{(2)}(t)$. The slope of the curve $Dv_{_M}^{(2)}(t)$ is
$0.532(5)$. Including a power-law correction, e.g., $Dv^{(2)}(t)
\sim t^b (1 + c / t )$, may improve the fitting to the numerical
data, and the resulting exponent is consistent with $0.532(5)$
within errors. According to Eq.~(\ref{equ230}), one can calculate
the exponent $\lambda_{_M} = 2.04(5)$ from $[1 - 2 \beta / \nu +
\lambda_{_M}(\zeta - 1)]/z = 0.532$. In Fig.~ \ref{f2}(b), the local
fluctuation $v_m^{(2)}(t)$ is plotted, and the slope of the curve is
$0.112(3)$. According to Eq.~(\ref{equ240}), $\lambda_{_M} =
3.06(3)$ can be derived from $[ - 2 \beta / \nu + \lambda_m(\zeta -
1)]/z = -0.112$. To study possible finite-size effects,
$v_{_M}^{(2)}(t)$ and $v_m^{(2)}(t)$ computed with different lattice
sizes at $H_c = 1.2933$ are also shown in Fig.~\ref{f2}. All the
curves in Fig.~\ref{f2}(a) are rescaled by a factor $L$, based on
Eq.~(\ref{equ230}). As it can be seen in the figure, the finite-size
effect can be easily controlled, i.e., it drops rapidly as $L$
increases. This is a merit of the short-time dynamic approach
\cite{zhe98,zhe99,luo98,zho09}.

\begin{figure}[ht]
\epsfysize=7.0cm \epsfclipoff \fboxsep=0pt
\setlength{\unitlength}{1.cm}
\begin{picture}(10,6)(0,0)
\put(-3.0,-0.3){{\epsffile{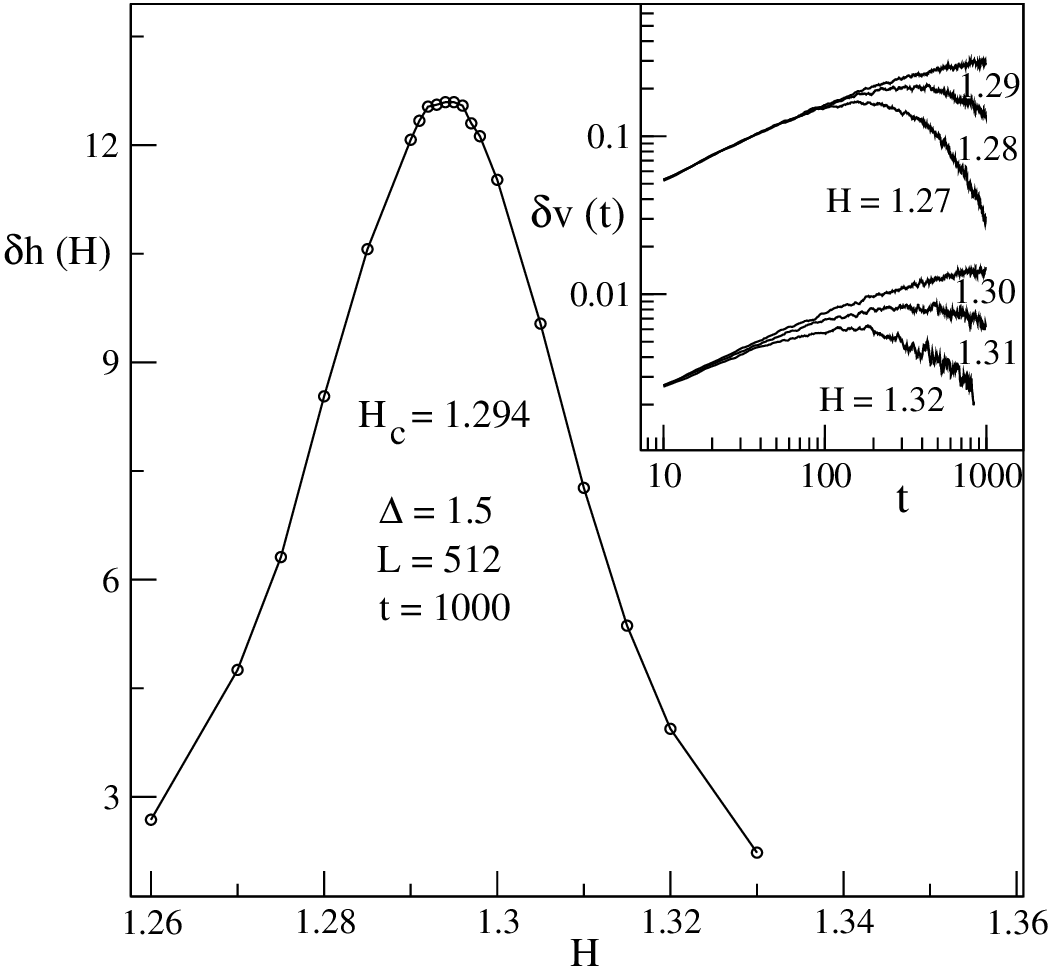}}}\epsfysize=7.0cm
\put(5.2,-0.3){{\epsffile{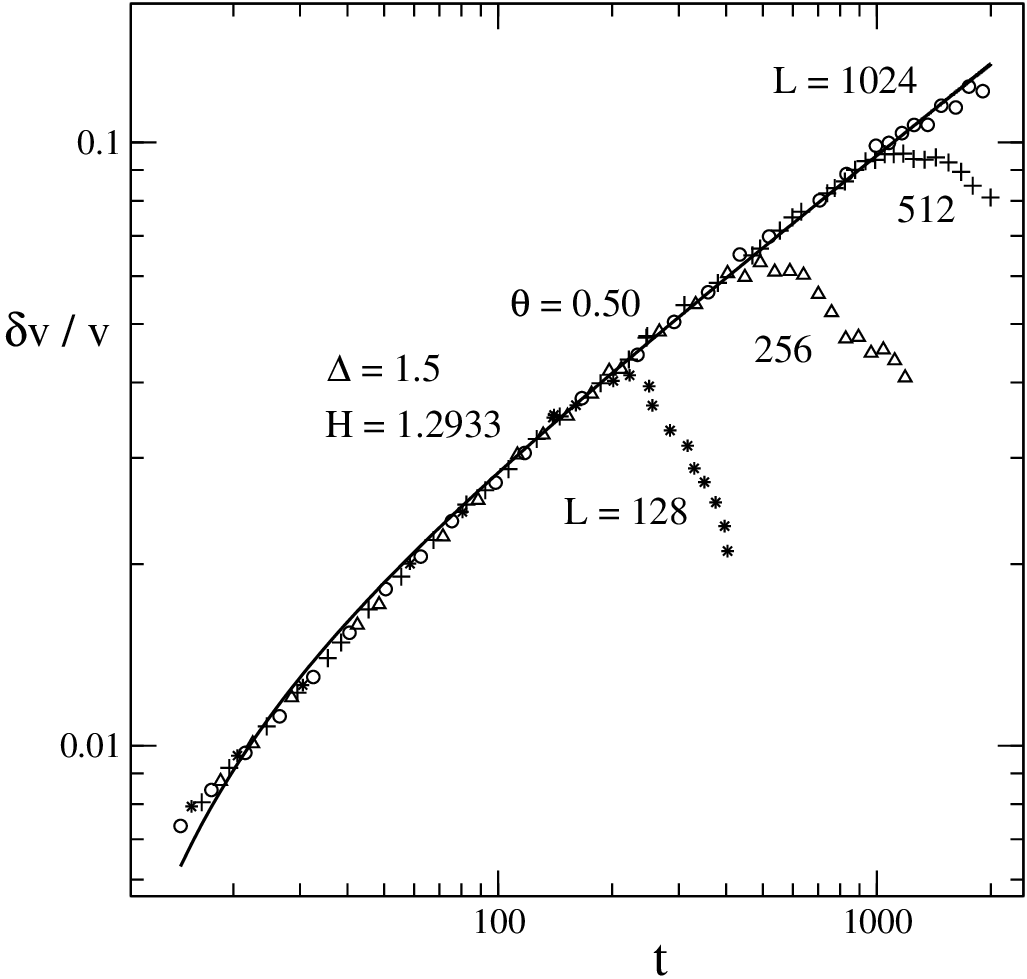}}}
\end{picture}

\hspace{-5.0cm}\footnotesize{(a)}\hspace{8.0cm}\footnotesize{(b)}
\caption{(a) The overhang size $\delta h(t,H)$ is plotted as a
function of the driving field $H$ at $t = 1000$. In the inset, the
overhang velocity $\delta v(t)$ is shown for different $H$ on a
log-log scale. For clarity, the curves for $H=1.29, 1.28$ and $1.27$
are shifted up. (b) $\delta v / v $ is displayed at $H_c = 1.2933$
for different $L$ on a log-log scale. The solid line represents a
power-law fit with correction.} \label{f3}
\end{figure}

However, both $Dv_{_E}^{(2)}(t)$ and $v_e^{(2)}(t)$ do not exhibit a
power-law behavior either at $H_c=1.2913$ or $1.2933$. In other
words, they do not obey the scaling forms in Eqs.~(\ref{equ230}) and
(\ref{equ240}). Additionally, $v_e(y,t)$ does not show a standard
correlation in the $y$ direction. In fact, the domain interface
defined with the envelop may be considered as adding overhangs and
islands to the domain interface defined with the magnetization. To
understand the domain interface defined with the envelop, we may
alternatively investigate the dynamics of overhangs and islands. In
the inset of Fig.~\ref{f3}(a), we present the overhang velocity
$\delta v(t)$ for different driving field $H$. Different from the
velocity $v_{_M}(t)$ and $v_{_E}(t)$ of the domain interface,
$\delta v(t)$ drops rapidly down for {\it both smaller and larger}
driving field $H$, and reaches maximum at the transition point
$H_c$, which is estimated to be between $1.29$ and $1.30$. In order
to determine the transition field more accurately, we calculate the
overhang size $\delta h(t,H)$ as a function of the external field
$H$ at $t=1000$. As shown in Fig.~\ref{f3}(a), the maximum of
$\delta h(H)$ yields $H_c = 1.294(1)$, in good agreement with the
transition field $H_c =1.2933(2)$ estimated from $v_{_M}(t)$. The
dynamic effect of overhangs and islands is the most prominent at the
depinning transition point. In Fig.~\ref{f3}(b), $\delta v / v$ is
plotted at $H=1.2933$ with different lattice size $L$. The
finite-size effect is negligible for $L=1024$ up to $t=2000$.
According to Eq.~(\ref{equ250}), $\theta = 0.50(2)$ is obtained with
a power-law fit for $t>100$, and a power-law correction improves the
fitting to the numerical data.

\begin{figure}[ht]
\epsfysize=7.0cm \epsfclipoff \fboxsep=0pt
\setlength{\unitlength}{1.cm}
\begin{picture}(10,6)(0,0)
\put(-3.0,-0.3){{\epsffile{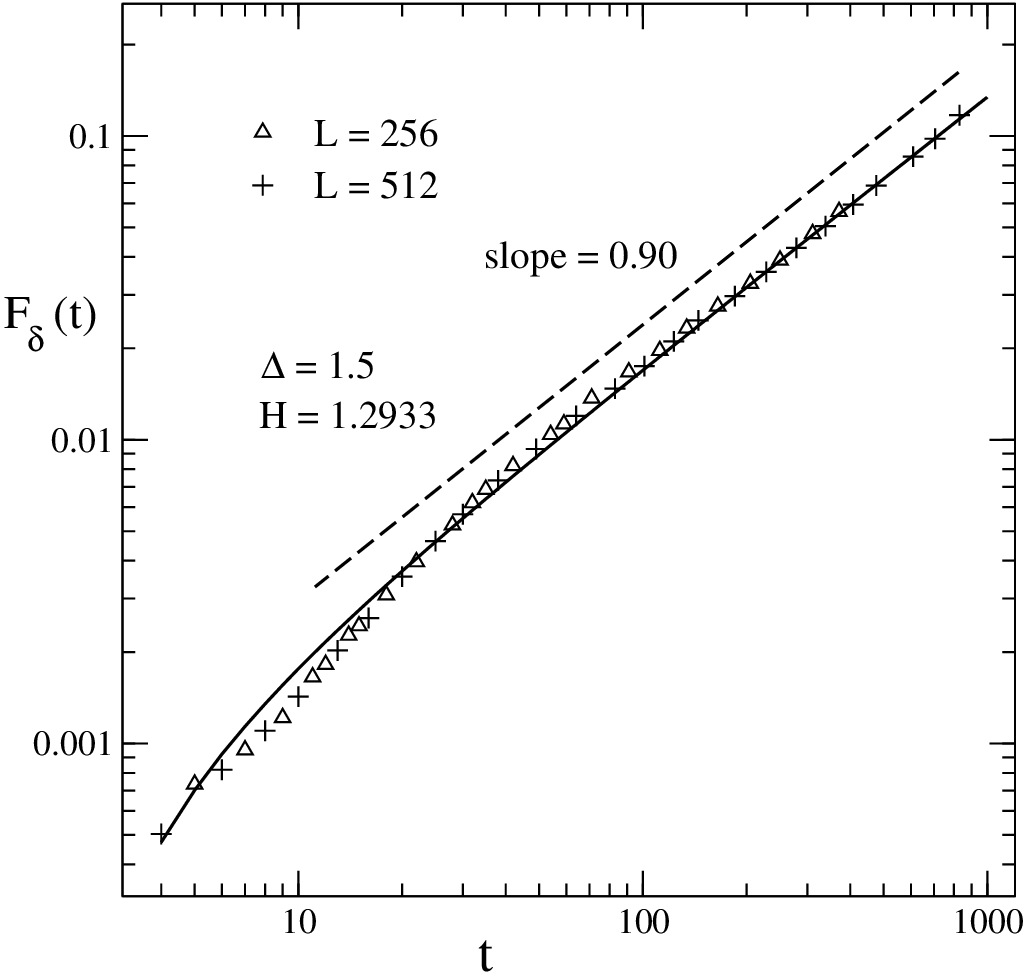}}}
\put(5.2,-0.3){{\epsffile{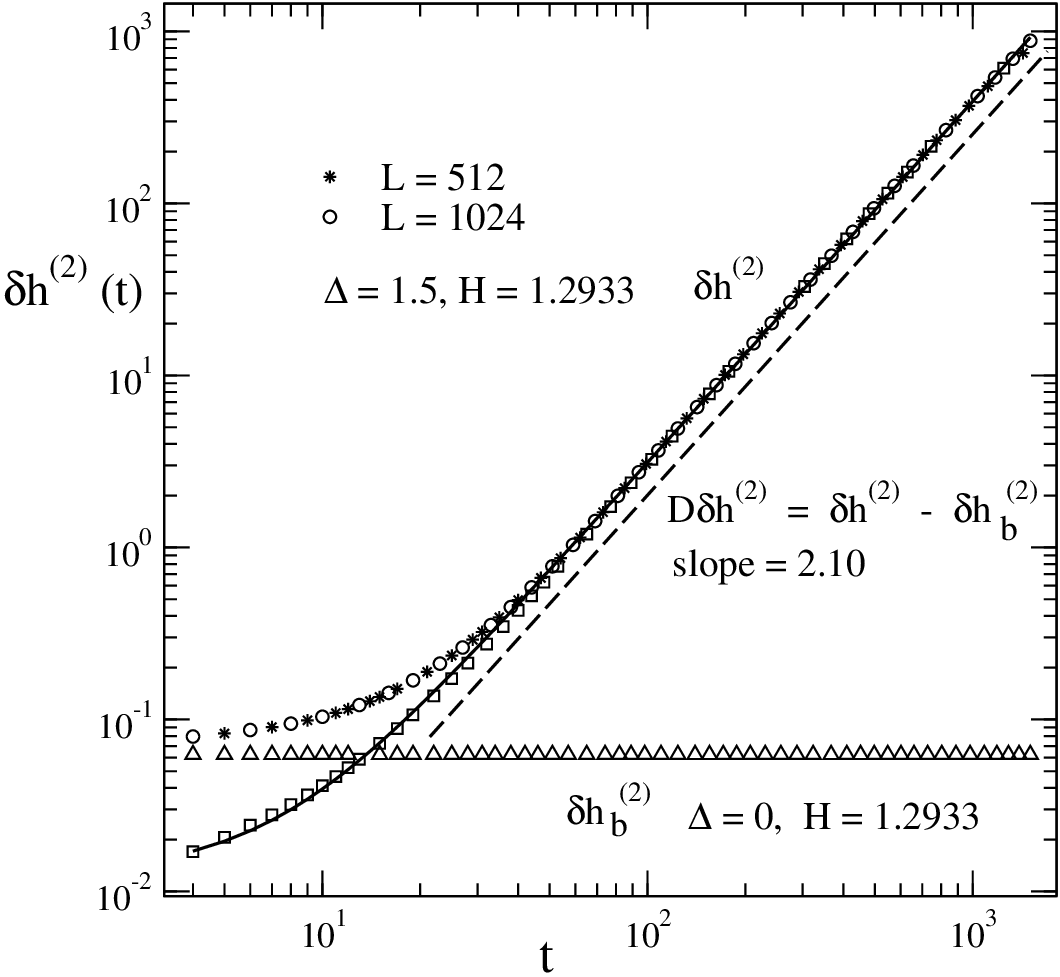}}}\epsfysize=7.0cm
\end{picture}

\hspace{-5.0cm}\footnotesize{(a)}\hspace{8.0cm}\footnotesize{(b)}
\caption{(a) $F_\delta(t)$ is plotted at $H_c=1.2933$. The curve of
$L=256$ has been rescaled by a factor $1/2$, according to
Eq.~(\ref{equ265}). (b) The roughness function and pure roughness
function of overhangs and islands are displayed at $H_c = 1.2933$.
In both (a) and (b), dashed lines show power-law fits, while solid
lines are for power-law fits with correction.}\label{f4}
\end{figure}

To determine the dynamic exponent $z_\delta$, we plot the function
$F_\delta(t)$ at $H_c =1.2933$ in Fig.~\ref{f4}(a). The slope of the
curve with $L=512$ is $0.90(1)$, and it yields $z_\delta=1.11(1)$.
An power-law correction may improve the fitting to the numerical
data. Collapse of the curves with $L=256$ and $512$ indicates the
$L$-dependence $F_\delta(t) \sim 1/L$. In Fig.~\ref{f4}(b), the
roughness functions $\delta h^{(2)}(t)$ and $\delta h_b^{(2)}(t)$
for $\Delta=1.5$ and $0$, and the pure roughness function $D\delta
h^{(2)}(t)$ are displayed at $H=1.2933$. Obviously, $D\delta
h^{(2)}(t)$ exhibits a cleaner power-law behavior than $\delta
h^{(2)}(t)$. The slop of the curve is $2.10(3) $, and it leads to
$\zeta_{\delta} = 1.16(2)$, close to $\zeta = 1.14$ for the domain
interface. With a correction to scaling, i.e., $D\delta
h^{(2)}(t)\sim t^{2\zeta_{\delta}/z_\delta}(1 + c/t^2)$, the fitting
to numerical data is extended. Since both the overhang size and
velocity reaches a maximum at the transition field $H_c$, the
standard exponent $\nu$ is not defined. In other words, $\partial_H
\delta v(t,H)$ is close to zero at $H_c$, and effectively,
$\nu_\delta \gg 1$.

\begin{figure}[ht]
\epsfysize=7.0cm \epsfclipoff \fboxsep=0pt
\setlength{\unitlength}{1.cm}
\begin{picture}(10,6)(0,0)
\put(1.0,-0.3){{\epsffile{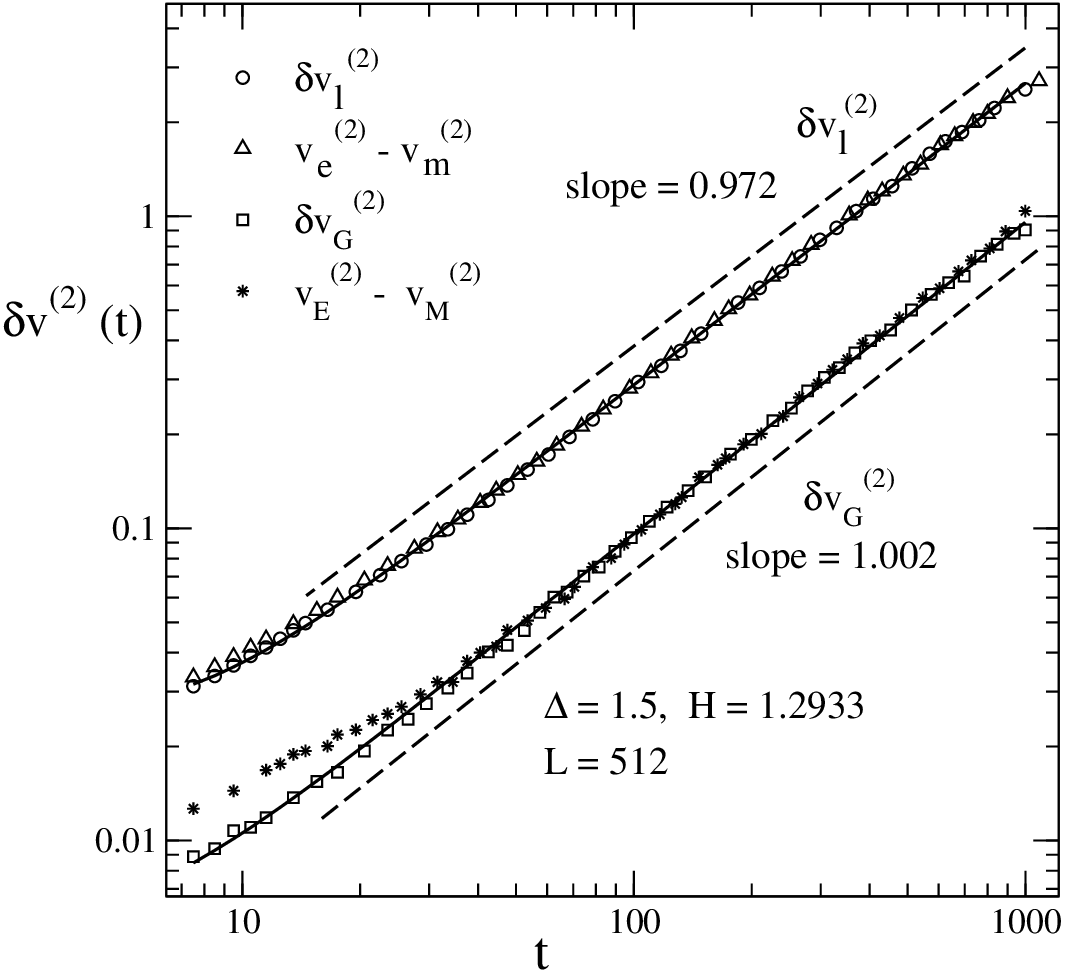}}}
\end{picture}

\caption{The local and global fluctuations of the overhang velocity,
$\delta v_l^{(2)}$ and $\delta v_G^{(2)}$, are plotted on a log-log
scale. For comparison, $v_e^{(2)} - v_m^{(2)}$ and $v_{_E}^{(2)} -
v_{_M}^{(2)}$ are also displayed. To show the finite-size
dependence, the global quantities are rescaled by a factor of $L/4$.
Dashed lines show power-law fits, while solid lines are for
power-law fits with correction.}\label{f5}
\end{figure}

Finally, we consider the local and global fluctuations of the
overhang velocity. In Fig.~\ref{f5}, the local fluctuation $\delta
v_l^{(2)}(t)$ is plotted at $H = 1.2933$. By the definition, the
local fluctuation can also be calculated with $\delta v_l^{(2)}(t) =
v_e^{(2)}(t) - v_m^{(2)}(t) - 2\Delta (t)$, and $\Delta (t) =<
v_m(y,t)\delta v(y,t)>- v_{_M}(t)\delta v(t)$. Hence
$v_e^{(2)}(t)-v_m^{(2)}(t)$ is plotted for comparison. The result
indicates that $\delta v_l^{(2)}(t) \approx v_e^{(2)}(t) -
v_m^{(2)}(t)$, i.e., $\Delta (t) \approx 0$. According to
Eq.~(\ref{equ270}), a direct measurement from the slope of the curve
gives $2\alpha = 0.972(4)$. A power-law correction to scaling yields
a similar result $2\alpha = 0.976$. The global fluctuation $\delta
v_G^{(2)}(t)$ and $v_{_E}^{(2)}(t)-v_{_M}^{(2)}(t)$ are also plotted
in Fig.~\ref{f5}. Both of them are rescaled by a factor of $L/4$,
because of the finite-size dependence of $L^{-1}$. Overlapping of
these two curves is observed and it yields $\Delta (t) \approx 0$,
too. The exponent $2\alpha = 1.002(5)$ is determined from the slope,
in agreement with that measured from the local fluctuation of the
velocity. Actually, our numerical simulations show that $\delta
v_l^{(2)}(t) \approx L\ \delta v_G^{(2)}(t)$. In Fig.~\ref{f5}, it
is only for clarity that $\delta v_G^{(2)}(t)$ is rescaled by a
factor of $L/4$ rather than $L$. The result $\Delta (t) \approx 0$
indicates that $v_m(y,t)$ and $\delta v(y,t)$ are not correlated,
and therefore, another set of critical exponents is needed for
describing the dynamic behavior of overhangs and islands.

The fact, that both $\delta v_l^{(2)}(t)$ and $\delta v_G^{(2)}(t)$
are governed by a same exponent $\alpha \approx 0.5$, indicates that
the overhang velocity $\delta v(y, t)$ is {\it not} correlated in
the $y$ direction, although the overhang size $\delta h(y, t)$ does.
If we consider $\delta v(y, t)$ as "a height function", its
roughening process described by $\delta v_l^{(2)}(t)$ belongs to the
universality class of random depositions. Why does such a phenomenon
occur? For example, $h_{e}(y+1,t)$ may suddenly produce an overhang
at $h_{e}(y,t)$, and induce a rapid increase of the overhang
velocity $\delta v(y,t)$. However, this is not correlated with
$\delta v(y+1,t)$. For the domain interface defined with the
envelop, $v_{e}(y,t)$ is also not correlated in the $y$ direction,
since $v_{e}(y,t)=v_{m}(y,t)+\delta v(y, t)$, and $\delta v(y, t)$
dominates at larger $t$. At the transition field $H_c$, our
numerical simulations show that $ v_e^{(2)}(t) \approx v_m^{(2)}(t)+
\delta v_l^{(2)}(t)$, and $ v_{_E}^{(2)}(t) \approx v_{_M}^{(2)}(t)+
\delta v_{_G}^{(2)}(t)$. On the other hand, $\delta v_l^{(2)}(t)$
and $\delta v_{_G}^{(2)}(t)$ exhibit different power-law behaviors
from $v_m^{(2)}(t)$ and $v_{_M}^{(2)}(t)$ respectively. Therefore,
$v_e^{(2)}(t)$ and $ v_{_E}^{(2)}(t)$ do not obey a simple power
law. For large $t$, however, $\delta v_l^{(2)}(t)$ and $\delta
v_{_G}^{(2)}(t)$ dominate the dynamic behaviors of $v_e^{(2)}(t)$
and $ v_{_E}^{(2)}(t)$.

\section{Summary}

Based on the short-time dynamic approach, we have systematically
investigated the domain-wall dynamics of the DRFIM model at the
depinning transition, and have accurately determined the transition
field and all the static and dynamic critical exponents. Through two
different definitions of the domain interface, we examine the
dynamics of overhangs and islands. All the critical exponents for
the domain interface and for overhangs and islands are summarized in
Table~\ref{t1}.

For the domain interface defined with the envelop, we do observe the
dynamic scaling behaviors in Eqs.~(\ref{equ170})-(\ref{equ205}) at
an effective transition field $H_c=1.2913$, slightly below
$H_c=1.2933$ determined from the interface velocity defined with the
magnetization. The difference of the critical exponents between the
DRFIM model and QEW equation becomes lager for the domain interface
defined with the envelop, especially for the local roughness
exponent $\zeta_{loc}$. These results further support that the DRFIM
model and QEW equation are not in a same universality class. Since
the dynamic effect of overhangs and islands is maximally expressed
in the domain interface defined with the envelop. We argue that it
is mainly the overhangs and islands that induces the difference
between the DRFIM model and QEW equation. Experiments report the
local roughness exponent $\zeta_{loc}<1$ \cite{lem98,jos98,met07}.

The global and local fluctuations of the interface velocity defined
with the magnetization exhibit the power-law scaling behaviors
described by the exponents $\lambda_{_M}$ and $\lambda_m$ in
Eqs.~(\ref{equ230}) and (\ref{equ240}), while those defined with the
envelop do not. It indicates that the interface velocity defined
with the envelop is not a good order parameter of the depinning
transition, although it is closer to experiments with imaging
techniques. In fact, $v_{e}(y,t)$ is also not correlated in the $y$
direction. It should be interesting to measure the fluctuation and
correlation of the interface velocity in experiments.

At the depinning transition, the dynamics effect of overhangs and
islands reaches maximum. The observables of overhangs and islands do
obey the dynamic scaling forms in
Eqs.~(\ref{equ250})-(\ref{equ270}), similar to those for the domain
interface. However, another set of critical exponents should be
introduced, since $v_m(y,t)$ and $\delta v(y,t)$ are not correlated.
Different from the interface velocity, the overhang velocity
increases with time. The dynamic exponent $z_\delta=1.11(1)$ for
overhangs and islands is smaller than $z_\delta=1.33(1)$ for the
domain interface, while the roughness exponent is about the same. In
particular, the overhang velocity $\delta v(y,t)$ does not show
correlation in the $y$ direction, although the overhang size $\delta
h(y,t)$ does. Considering $\delta v(y,t)$ as a height function, its
roughening process belongs to the universality class of random
depositions. Since $h_{e}(y,t)=h_{m}(y,t)+\delta h(y, t)$, the
domain interface defined with the envelop can be mostly understood
by adding overhangs and islands to that defined with the
magnetization.

{\bf Acknowledgements:} This work was supported in part by NNSF of
China under grant No. 10875102, and Zhejiang Provincial Natural
Science Foundation of China under grant No. Z6090130.


\newpage

\begin{table}[h]\centering
\begin{tabular}[t]{r c c c| c c}
\hline
\hline  &   &  QEW & Magnetization  & Envelop & Overhang\\
\hline
$v(t)$        &   $H_c$            &                         &      1.2933(2)      &     1.2913(4)   &  1.294(1)                        \\
              &   $\beta$          & 0.33(2); 0.33           &      0.295(3)       &     0.278(4)    &             \\
              &   $\nu$            & 1.29(5); 1.33; 1.33(1)  &      1.02(2)        &     1.02(4)     &   $\gg 1$ \\
              &   $z $             & 1.5; 1.53               &      1.33(1)        &     1.28(1)     &  1.11(1)              \\
              &  $\theta$  && &&0.50(2)\\
$\omega^2(t)$ &   $\zeta$          & 1.26(1); 1.25; 1.24     &      1.14(1)        &     1.14(1)     &  1.16(2)       \\
$C(r,t)$      &   $\zeta$          & 1.23(1); 1.25           &      1.13(1)        &     1.14(1)     &            \\
              &   $\zeta_{loc}$    & 0.98; 0.92              &      0.735(8)       &     0.569(6)    &                 \\
\hline
$v_{_M}^{(2)}(t)$ & $\lambda$ && 2.04(5)&&\\

$v_{m}^{(2)}(t)$ &  && 3.06(3)&&\\
$\delta v_{_G}^{(2)}(t)$                  & $\alpha$ && &&0.501(3)\\
$\delta v_{l}^{(2)}(t)$                  &  && &&0.488(4)\\
\hline \hline
\end{tabular}
\caption{The depinning transition field and critical exponents
obtained for the DRFIM model are compared with those for the QEW
equation in Refs.~\cite{due05,kol06,kol06a,lop97,ros03}. The
exponents $\beta$, $\nu$, $z$, $\zeta$ and $\zeta_{loc}$ for the
domain interface defined with the magnetization are taken from
Ref.~\cite{zho09}.} \label{t1}
\end{table}

\end{document}